\documentstyle[11pt,aaspp4,flushrt]{article}
\def \beq{\begin{equation}}
\def \eeq{\end{equation}}
\def \beqa{\begin{eqnarray}}
\def \eeqa{\end{eqnarray}}
\begin{document}

\title{High-Redshift Galaxies: Their Predicted Size and Surface
Brightness Distributions and Their Gravitational Lensing Probability}

\author{Rennan Barkana\footnote{email: barkana@ias.edu}}
\affil{Institute for Advanced Study, Olden Lane, Princeton,
NJ 08540}

\author{Abraham Loeb\footnote{email: aloeb@cfa.harvard.edu}}
\affil{Astronomy Department, Harvard University, 60 Garden 
St., Cambridge, MA 02138}

\begin{abstract}

Direct observations of the first generation of luminous objects will
likely become feasible over the next decade. The advent of the Next
Generation Space Telescope ({\it NGST}\,) will allow imaging of
numerous galaxies and mini-quasars at redshifts $z \ga 5$. We apply
semi-analytic models of structure formation to estimate the rate of
multiple imaging of these sources by intervening gravitational
lenses. Popular CDM models for galaxy formation yield a lensing
optical depth of $\sim 1\%$ for sources at $z\sim 10$.  The expected
slope of the luminosity function of the early sources implies an
additional magnification bias of $\sim 5$, bringing the fraction of
lensed sources at $z=10$ to $\sim 5\%$.  We estimate the angular size
distribution of high-redshift disk galaxies and find that most of them
are more extended than the resolution limit of {\it NGST},\, $\sim
0\farcs06$.  We also show that there is only a modest redshift
evolution in the observed mean surface brightness of galaxies at $z\ga
2$. The expected increase by 1--2 orders of magnitude in the number of
resolved sources on the sky, due to observations with {\it NGST},\,
will dramatically improve upon the statistical significance of
existing weak lensing measurements. We show that, despite this
increase in the density of sources, confusion noise from $z>2$
galaxies is expected to be small for {\it NGST}\, observations.

\end{abstract}

\keywords{cosmology: theory --- gravitational lensing --- galaxies:
formation}

\section{Introduction}

Current observations reveal the existence of galaxies out to redshifts as
high as $z\sim 6.7$ (Chen et al.\ 1999; Weymann et al.\ 1998; Dey et al.\
1998; Spinrad et al.\ 1998; Hu, Cowie, \& McMahon 1998) and bright quasars
out to $z\sim 5$ (Fan et al.\ 1999). Based on sources for which high
resolution spectra are available, the intergalactic medium appears to be
predominantly ionized at this epoch, implying the existence of ionizing
sources at even higher redshifts (Madau 1999; Madau, Haardt, \& Rees 1999;
Haiman \& Loeb 1998a,b; Gnedin \& Ostriker 1997).

Hierarchical Cold dark matter (CDM) models for structure formation
predict that the first baryonic objects appeared near the Jeans mass
($\sim 10^5~{\rm M_\odot}$) at redshifts as high as $z\sim30$ (Haiman
\& Loeb 1998b, and references therein).  The Next Generation Space
Telescope ({\it NGST}\,), planned for launch in 2008, is expected to
reach an imaging sensitivity better than 1 nJy in the infrared, which
will allow it to detect galaxies or mini-quasars at $z\ga 10$.

In this paper we explore the ability of {\it NGST}\, to extend
gravitational lensing studies well beyond their current limits. Due to
the increased path length along the line-of-sight to the most distant
sources, their probability for being lensed is expected to be the
highest among all possible sources. Sources at $z>10$ will often be
lensed by $z>2$ galaxies, whose masses can then be determined with
lens modeling.  Similarly, the shape distortions (or weak lensing)
caused by foreground clusters of galaxies will be used to determine
the mass distributions of less massive and higher redshift clusters
than currently feasible. In addition to studying the lensing objects,
observers will exploit the magnification of the sources to resolve and
study more distant galaxies than would otherwise be possible.

The probability for strong gravitational lensing depends on the
abundance of lenses, their mass profiles, and the angular diameter
distances among the source, the lens and the observer. The statistics
of existing lens surveys have been used at low redshifts to constrain
the cosmological constant (for the most detailed work see Kochanek
1996a, and references therein), although substantial uncertainties
remain regarding the luminosity function of early-type galaxies and
their dark matter content (Cheng \& Krauss 1999; Chiba \& Yoshii
1999). The properties of dark matter halos will be better probed in
the future by individual as well as statistical studies of the large
samples of lenses expected from quasar surveys such as the 2-Degree
Field (Croom et al.\ 1998) and the Sloan Digital Sky Survey (SDSS
Collaboration 1996). Given the early stage of observations of the
redshift evolution of galaxies and their dark halos, we adopt a
theoretical approach in our analysis and use the abundance of dark
matter halos as predicted by the Press-Schechter (1974, hereafter PS)
model. A similar approach has been used previously for calculating
lensing statistics at low redshifts, with an emphasis on lenses with
image separations above $5\arcsec$ (Narayan \& White 1988; Kochanek
1995; Nakamura \& Suto 1997) or on lensing rates of supernovae
(Porciani \& Madau 1999).

Even when multiple images are not produced, the shape distortions
caused by weak lensing can be used to determine the lensing mass
distribution. Large numbers of sources are required in order to
average away the noise due to the intrinsic ellipticities of sources, and
so the mass distribution can only be determined for the extended halos
of rich clusters of galaxies (e.g., Hoekstra et al.\ 1998; Luppino \&
Kaiser 1997; Seitz et al.\ 1996) or statistically for galaxies (e.g.,
Brainerd et al.\ 1996; Hudson et al.\ 1998). Schneider \& Kneib (1998)
have noted that the ability of {\it NGST}\, to take deeper exposures
than is possible with current instruments will increase the observed
density of sources on the sky, particularly of those at high
redshifts. The large increase might allow such applications as
a detailed weak lensing mapping of substructure in clusters.
Obviously, the source galaxies must be well resolved to
allow an accurate shape measurement. Unfortunately, the characteristic
galaxy size is expected to decrease with redshift for two reasons: (i)
the mean density of collapsed objects scales as the density of the
Universe at the collapse redshift, namely as $(1+z)^3$. Hence, objects
of a given mass are expected to be more compact at high redshifts, and
(ii) the characteristic mass of collapsed objects decreases with
increasing redshift in the bottom-up CDM models of structure
formation. In the following, we attempt to calculate the size
distribution of high redshift sources. Aside from the obvious
implications for weak lensing studies, the finite size of sources also
has important implications for their detectability with {\it NGST}\,
above the background noise of the sky brightness.

The outline of the paper is as follows. In \S 2 we employ the PS halo
abundance in several hierarchical models of structure formation to
estimate the lensing rate of the high redshift objects that will be
observed with {\it NGST}.\, This lensing rate has been calculated by
Marri \& Ferrara (1998) assuming point mass lenses. We use the simple
but more realistic model of a singular isothermal sphere (SIS) profile
for dark matter halos and obtain a substantially lower lensing rate.
The formation of galactic disks and the distributions of their various
properties have been studied by Dalcanton, Spergel, \& Summers (1997)
and Mo, Mao, \& White (1998) in the framework of hierarchical models
of structure formation. In \S 3 we apply their models to high redshift
sources, and find the angular size distribution of galactic disks as a
function of redshift. We use this distribution to predict whether
observations with {\it NGST}\, will be significantly limited by
confusion noise. We also calculate the redshift evolution of the mean
surface brightness of disks. Finally, \S 4 summarizes the implications
of our results.

\section{Lensing Rate of High-Redshift Sources}

\subsection{Calculation Method}
\label{Vc}

We calculate the abundance of lenses based on the PS halo mass
function. Relevant expressions for various CDM cosmologies are given,
e.g., in Navarro, Frenk, \& White (1997, hereafter NFW).  The PS
abundance agrees with N-body simulations on the mass scale of galaxy
clusters, but may over-predict the abundance of galaxy halos at present
by a factor of 1.5--2 (e.g., Gross et al.\ 1998). At higher redshifts,
the characteristic mass scale of collapsed objects drops and the PS
abundance becomes more accurate for the galaxy-size halos which
dominate the lensing rate.

The probability for producing multiple images of a source at a
redshift $z_S$ due to gravitational lensing by SIS lenses is obtained
by integrating over lens redshift $z_L$ the differential optical depth
(Turner, Ostriker, \& Gott 1984; Fukugita et al.\ 1992; Peebles 1993)
\beq d\tau=16 \pi^3 n \left(\frac{\sigma}{c}\right)^4 (1+z_L)^3
\left(\frac{D_{OL} D_{LS}}{D_{OS}}\right)^2 \frac{c dt}{dz_L} d z_L\ ,
\label{dtau}
\eeq in terms of the comoving density of lenses $n$, velocity
dispersion $\sigma$, look-back time $t$, and angular diameter
distances $D$ among the observer, lens and source. More generally we
replace $n \sigma^4$ by \beq 
\label{nsig}
\langle n \sigma^4\rangle = \int
\frac{dn(M,z_L)}{dM} \sigma^4(M,z_L)\, dM\ , \eeq where $dn/dM$ is the
PS halo mass function. We assume that $\sigma(M,z)=V_c(M,z)/\sqrt{2}$\,,
and we calculate the circular velocity $V_c(M,z)$ corresponding to a halo
of a given mass as in NFW, except that we vary the virialization
overdensity using the fitting formula of Bryan \& Norman (1998). The
lensing rate depends on a combination of redshift factors, as well as
the evolution of halo abundance.  At higher redshifts, halos of a
given mass are more concentrated and have a higher $\sigma$, but
lower-mass halos contain most of the mass in the Universe.

When calculating the angular diameter distances we assume the standard
distance formulas in a homogeneous universe. Inhomogeneities, however,
cause a dispersion around the mean distance. The non-Gaussian, skewed
distribution of distances in hierarchical models is best studied with
numerical simulations (e.g., Wambsganss et al.\ 1998), and can in principle
be included self-consistently in more elaborate calculations of the lensing
statistics.

We consider cosmological models with various values of the cosmological
density parameters of matter and vacuum (cosmological constant), $\Omega_0$
and $\Omega_{\Lambda}$. In particular, we show results for $\Lambda$CDM
(with $\Omega_0=0.3$ and $\Omega_{\Lambda}=0.7$), OCDM (with $\Omega_0=0.3$
and $\Omega_{\Lambda}=0$), and SCDM (with $\Omega_0=1$ and
$\Omega_{\Lambda}=0$). The models assume a Hubble constant $h=0.5$ if
$\Omega_0=1$ and $h=0.7$ otherwise (where $H_0=100\, h\mbox{ km
s}^{-1}\mbox{Mpc}^{-1}$). They also assume a primordial scale invariant
($n=1$) power spectrum, normalized to the present cluster abundance,
$\sigma_8=0.5\ \Omega_0^{-0.5}$ (e.g., Pen 1998 and references therein),
where $\sigma_8$ is the root-mean-square amplitude of mass fluctuations in
spheres of radius $8\ h^{-1}$ Mpc.

\subsection{Numerical Results}

\begin{figure}
\epsscale{0.7}
\plotone{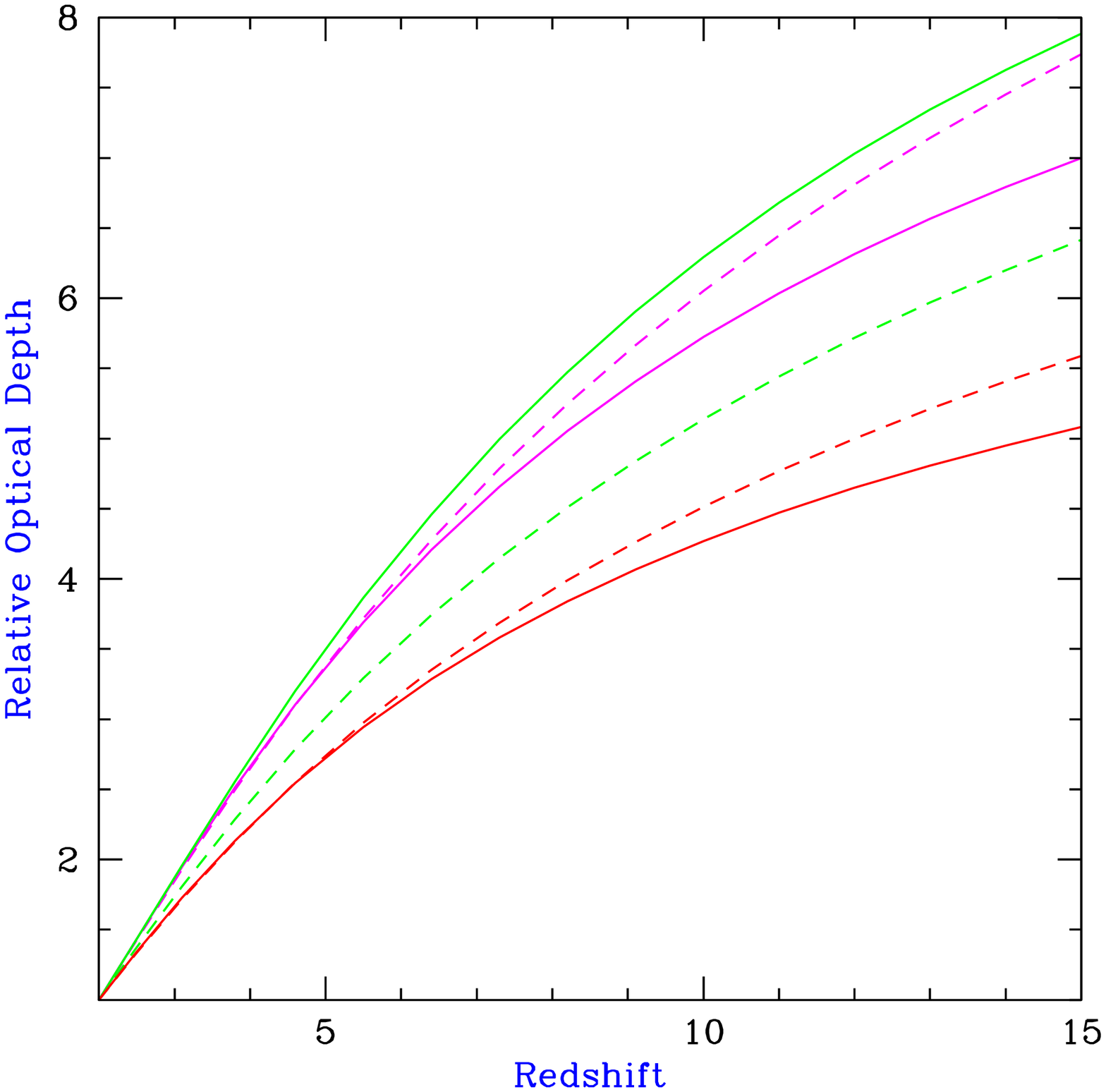}
\caption{Relative variation of the lensing optical depth with source
redshift (magnification bias not included), with each curve normalized
to unity at $z_S=2$. We plot the results for PS halos (solid curves) in
OCDM, $\Lambda$CDM, and SCDM, in order from top to bottom. These are
compared to the corresponding results for non-evolving lenses (dashed
curves) in $\Lambda$CDM, OCDM, and SCDM, in order from top to bottom.}
\label{figtau}
\end{figure}

In Figure \ref{figtau} we show the variation of the lensing optical
depth with source redshift. This plot does not include the
magnification bias which we discuss below. In order to show the
relative variation, we normalize each curve to unity at $z_S=2$. The
dashed curves show results for non-evolving lenses with $\langle n
\sigma^4 \rangle =const$ in $\Lambda$CDM, OCDM, and SCDM, in order
from top to bottom. The higher values obtained in low-$\Omega$ models
are due to the increased spatial volume in these cosmologies.  The
solid curves show the results for the PS halo distribution in OCDM,
$\Lambda$CDM, and SCDM, in order from top to bottom. High redshifts
are characterized by a decrease in $dn/dM$ at high masses and an
increase at low masses, so that the typical mass of collapsing objects
decreases.  In the OCDM model, the evolution of $dn/dM$ toward lower
masses is slow enough that $\langle n \sigma^4\rangle$ increases with
$z$ up to $z\sim3.5$, which increases the lensing optical depth above
the value expected for non-evolving lenses.

\begin{figure}
\epsscale{0.7}
\plotone{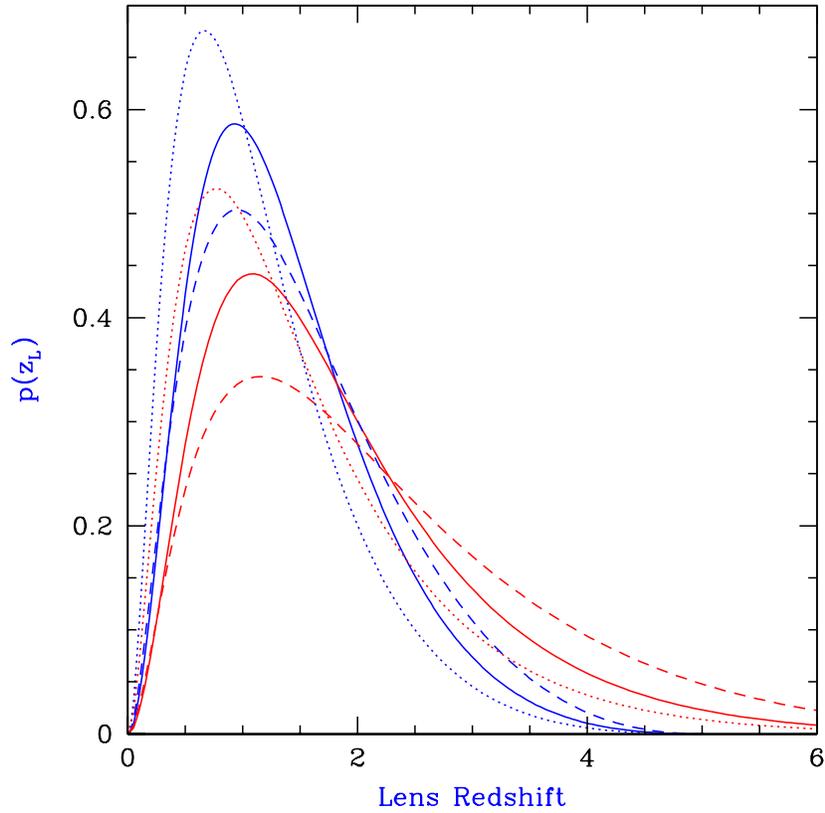}
\caption{Distribution of lens redshifts for a fixed source redshift,
for PS halos in $\Lambda$CDM (solid curves), OCDM (dashed curves), and
SCDM (dotted curves). In each model, curves are shown for a source at
$z_S=5$ and for $z_S=10$, where the higher curve at $z_L<1$
corresponds to $z_S=5$. The probability density $p(z_L)$ is shown,
where the fraction of lenses between $z_L$ and $z_L+dz_L$ is
$p(z_L)dz_L$.}
\label{figzLens}
\end{figure}

For a given source redshift, the distribution of lens redshifts is
proportional to $d\tau/dz_L$, which is given by eqs.\ (\ref{dtau}) and
(\ref{nsig}). In Figure \ref{figzLens} we show the probability density
$p(z_L)$, defined so that the fraction of lenses between $z_L$ and
$z_L+dz_L$ is $p(z_L)dz_L$. We assume PS halos in $\Lambda$CDM (solid
curves), OCDM (dashed curves), or SCDM (dotted curves). In each
cosmological model, we consider a source at $z_S=5$ or at $z_S=10$,
where the higher curve at $z_L<1$ corresponds to $z_S=5$. The curves
peak around $z_L=1$ in the low-$\Omega$ models and around $z_L=0.7$ in
SCDM. In each case a significant fraction of the lenses are above
redshift 2: $20\%$ for $z_S=5$ and $36\%$ for $z_S=10$ in
$\Lambda$CDM. The $z_L>2$ fractions are higher in OCDM ($26\%$ for
$z_S=5$ and $48\%$ for $z_S=10$) and lower in SCDM ($13\%$ for $z_S=5$
and $26\%$ for $z_S=10$). 

The fraction of lensed sources in an actual survey is enhanced,
relative to the above lensing probability, by the so-called
magnification bias. At a given observed flux level, unlensed sources
compete with lensed sources that are intrinsically fainter. Since
fainter galaxies are more numerous, the fraction of lenses in an
observed sample is larger than the optical depth discussed above. The
magnification bias is calculated in detail below, but for the purpose
of the discussion here we adopt a uniform enhancement factor of 5 when
computing the lensing fraction.  Our results for the different
cosmological models are summarized in Table 1.  At $z_S=2$ we compare
the results from the hierarchical PS models to a no-evolution model of
the lens population based on the local luminosity function of
galaxies. The last column of Table 1 shows the results (with a
magnification bias factor of 5), for example, for the parameters of
the no-evolution model of Kochanek (1996a), who adopted a number
density $n_e=6.1\ h^3\times 10^{-3}\ $Mpc$^{-3}$ of E/S0 galaxies, a
Schechter function slope $\alpha=-1$, a Faber-Jackson exponent
$\gamma=4$, and a characteristic dark matter velocity dispersion
$\sigma_{\star} =225~{\rm km~s^{-1}}$. The PS models yield a higher
lensing fraction, although the difference is small for the
$\Lambda$CDM model.  In all the PS models, the fraction of multiply
imaged systems at $z_S=10$ is around $5\%$ if the magnification bias
is 5.

\begin{figure}
\epsscale{0.7}
\plotone{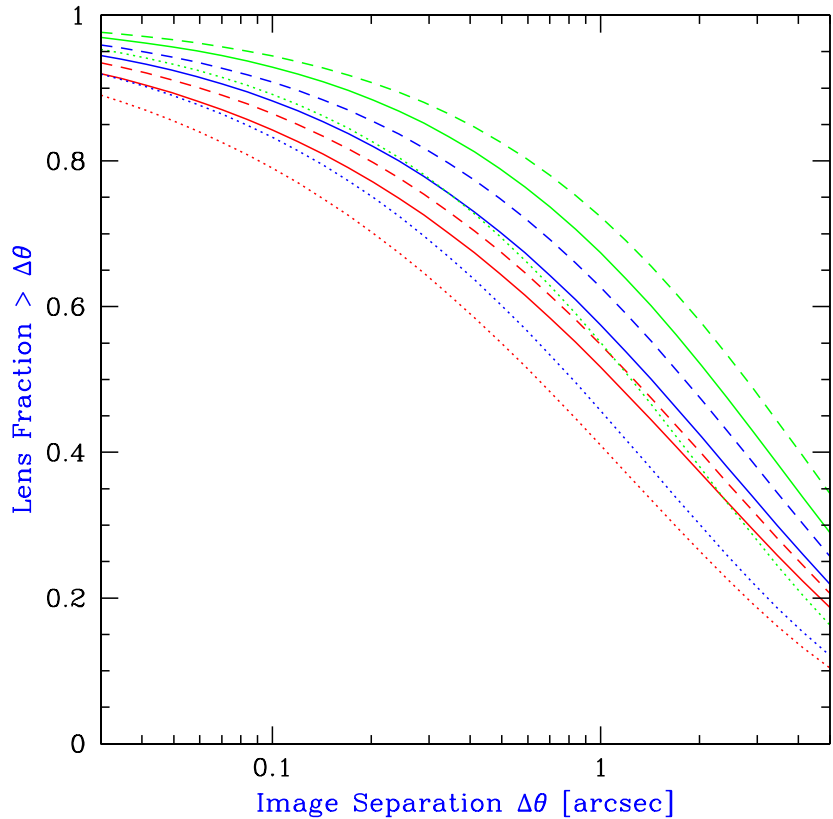}
\caption{Integral distribution of image separation, for PS halos
in $\Lambda$CDM (solid curves), OCDM (dashed curves), and SCDM
(dotted curves). The results in each cosmological model are shown
for sources at $z_S=2$, 5, and 10, from top to bottom.}
\label{figang}
\end{figure}

In the SIS model, the two images of a multiply-imaged source have a
fixed angular separation, independent of source position, of $\Delta
\theta=8\pi (\sigma/c)^2 (D_{LS}/D_{OS})$. The overall distribution of
angular separations is shown in Figure \ref{figang} for $\Lambda$CDM
(solid curves), OCDM (dashed curves), and SCDM (dotted curves). The
results are illustrated for $z_S=2$, 5, and 10 in each model. Image
separations are typically reduced by a factor of 2--3 between $z_S=2$
and $z_S=10$, almost entirely due to the evolution of the lenses.
With the {\it NGST}\, resolution of $\sim 0\farcs06$, a large majority
($\sim 85\%$) of lenses with $\Delta \theta < 5 \arcsec$ can be
resolved even for $z_S=10$. Note, however, that a ground-based survey
with $\sim 1 \arcsec$ seeing is likely to miss $\sim 60\%$ of these
lenses. There is also a tail of lenses with separations $\Delta \theta
> 5 \arcsec$. These large separation lenses, and the observational
difficulties in identifying them, have been previously explored both
analytically (Narayan \& White 1988; Kochanek 1995) and with numerical
simulations (Cen et al.\ 1994; Wambsganss et al.\ 1995).

The magnification bias is determined by the distribution of image
magnifications and by the source luminosity function.  Denoting the
probability distribution of magnifications by $q(A)$ (independent of $z_L$
and $z_S$ for the SIS), and the number counts of sources per unit flux at a
flux $F$ by $dN/dF$, the fraction of lensed sources at the observed flux
$F$ is increased by a bias factor \beq B=\int
\left.\frac{dN}{dF}\right|_{F/A} \left[\left.
\frac{dN}{dF}\right|_F\right]^{-1}\,q(A)\frac{dA}{A}\ .\eeq As noted above,
{\it NGST}\, will resolve almost all double images and so we count them as
two apparent sources. Thus we compute the bias factors separately for the
two images, using $q(A)=2/(A-1)^3$ and $A>2$ for the brighter image, and
$q(A)=2/(A+1)^3$ and $A>0$ for the fainter image. 
We then find the sum, which is dominated by the brighter image of the
two. This sum includes the contributions to sources observed at a flux $F$
from all lensed images (each of which is either the bright image or the faint
image of a lensed pair). The product of the resulting bias factor and the
lensing optical depth yields the fraction of all apparent sources which are
part of a lensed system. We note that any attempt to estimate the
magnification bias of high-redshift sources is highly uncertain at the
present due to several tentative assumptions about their characteristic
mass-to-light ratio, star formation history, initial stellar mass function,
dust extinction amplitude, and quasar formation history.

\begin{figure}
\epsscale{0.7}
\plotone{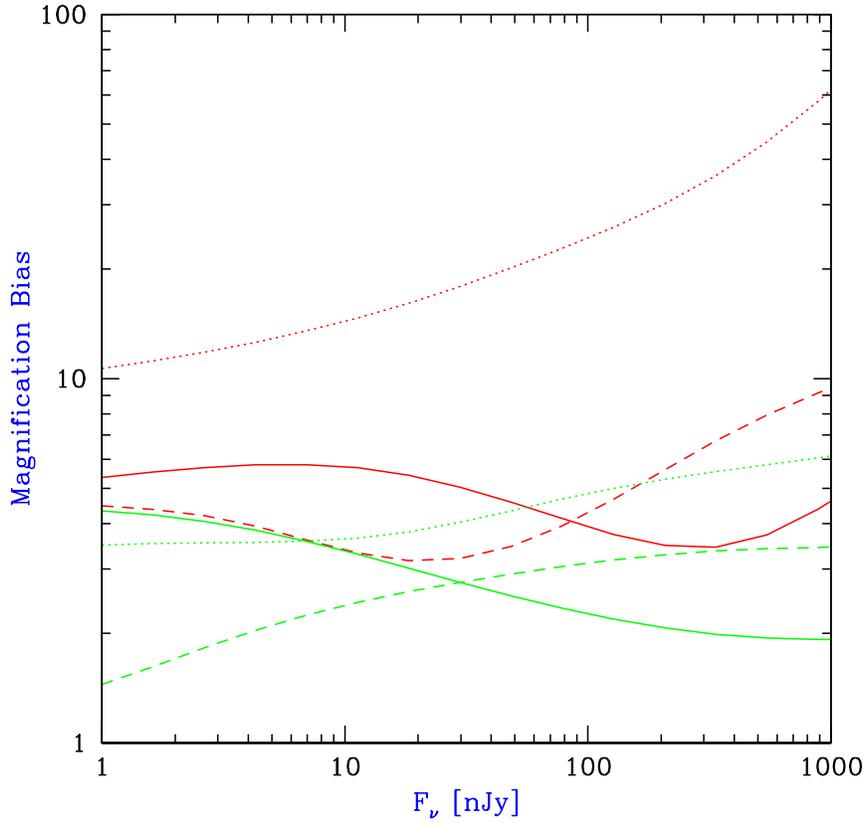}
\caption{Magnification bias for the Haiman \& Loeb (1998b) number
counts in $\Lambda$CDM (extrapolated as power laws to fluxes fainter
than 0.1 nJy). Solid lines are for mini-quasars, dashed lines are for
galaxies undergoing starbursts which convert $20\%$ of the gas in each
halo into stars, and dotted lines are for starbursts that use only
$2\%$ of the gas in each halo. For each type of source, the lower
curve is for all sources at $z_S>5$ and the higher one is for all
sources at $z_S>10$.  }
\label{figbias}
\end{figure}

Figure \ref{figbias} illustrates the magnification bias for the {\it
NGST}\, number count model of Haiman \& Loeb (1998b; 1997), who
assumed cosmological parameters nearly equivalent to our $\Lambda$CDM
model. Solid lines are for mini-quasars, dashed lines are for galaxies
undergoing starbursts which convert $20\%$ of the gas of each halo
into stars, and dotted lines are for starbursts which use only $2\%$
of the gas of each halo. For each type of source, we show separate
curves corresponding to all sources at redshifts $z_S>5$ or to all
sources at redshifts $z_S>10$. Although the $z_S>10$ number counts are
smaller, they are steeper than the $z_S>5$ counts and produce a larger
magnification bias. Similarly, for low $(2\%)$ star-formation
efficiency, galaxies are detected only if they lie in relatively
massive halos, which have a steeper mass function and thus a larger
magnification bias than for a higher star-formation efficiency. These
results indicate a magnification bias around 3--6, but this factor
could be much higher if the actual number counts are only somewhat
steeper than predicted by these models. Indeed, the number counts fall
off roughly as power laws $dN/dF_{\nu} \propto F_{\nu}^{-\beta}$ with
$\beta \sim$ 2--2.5, while for the SIS, the magnification bias
diverges at the critical value $\beta=3$. Using Figure \ref{figbias},
Table 1, and the number counts of Haiman \& Loeb (1998b), the
estimated number of sources (lensed sources) above 1 nJy per $4
\arcmin \times 4 \arcmin$ field of view is 90 (5) for $z>10$ quasars,
300 (12) for $z>5$ quasars, 400 (17) for $z>10$ galaxies with $20\%$
star-formation efficiency, $10^4$ (200) for $z>5$ galaxies with $20\%$
efficiency, 20 (1) for $z>10$ galaxies with $2\%$ efficiency, and
$2\times 10^3$ (30) for $z>5$ galaxies with $2\%$ efficiency. Note,
however, that the number counts for galaxies are reduced when we
include the fact that most galaxies are resolved by {\it NGST}\, and
cannot be treated as point sources (see \S 3).

We have assumed that each lensing halo can be approximated as a SIS,
although the mass distributions in actual halos might be more
complicated. Numerical simulations of pure dark matter indicate a
roughly universal profile (NFW) with a $1/r$ density profile in the
core. This result is supported by very high resolution simulations of
a small number of halos (Moore et al.\ 1999), although simulations of
large numbers of halos typically find a shallower inner density
profile in agreement with observed rotation curves of
dark-matter-dominated galaxies (Kravtsov et al.\ 1998). In addition,
galaxy halos undergo adiabatic compression when the baryons cool and
contract (e.g., Flores et al.\ 1993). Halos with the NFW profile have
a smaller lensing cross-section than the SIS, but this is partly
compensated for by the higher mean magnification and thus the higher
magnification bias produced by NFW lenses (Keeton 1999, in
preparation). In the above discussion, we have also assumed spherical
symmetry. If the SIS is made ellipsoidal, with an ellipticity of 0.3,
then the total lensing cross-section is changed only slightly, but
lenses above a total magnification of $\sim 8$ are then mostly
four-image systems (see, e.g., Kochanek 1996b).  We have also assumed
that each halo acts as an isolated lens, while in reality galaxies are
clustered and many galaxies lie in small groups.  The large dark
matter halo associated with the group may combine with the halos of
the individual galaxies and enhance their lensing
cross-section. External shear due to group halos will also tend to
increase the fraction of four-image systems. On the other hand, dust
extinction may reduce the number of lensed systems below our
estimates, especially since high redshift galaxies are observed at
rest-frame UV wavelengths. Significant extinction may arise from dust
in the source galaxy itself as well as dust in the lens galaxy, if the
image path passes sufficiently close to the center of the lens galaxy.

\section{Size Distribution of High-Redshift Disk Galaxies}

\subsection{Semi-Analytic Model}
\label{limits}

The formation of disk galaxies within hierarchical models of structure
formation was explored by Fall \& Efstathiou (1980).  More recently,
the distribution of disk sizes was derived and compared to
observations by Dalcanton, Spergel, \& Summers (1997; hereafter DSS)
and Mo, Mao, \& White (1998; hereafter MMW). In order to estimate the
ability of {\it NGST}\, to resolve high redshift disks, we adopt the
simple model of an exponential disk in a SIS halo. We consider a halo
of mass $M$, virial radius $r_{\rm vir}$, total energy $E$, and
angular momentum $J$, for which the spin parameter is defined as \beq
\lambda \equiv J |E|^{1/2} G^{-1} M^{-5/2}\ . \eeq If the disk mass is
a fraction $m_d$ of the halo mass and its angular momentum is a
fraction $j_d$ of that of the halo, then the exponential scale radius
of the disk is given by (MMW) \beq
R_d=\frac{1}{\sqrt{2}}\left(\frac{j_d}{m_d}\right) \lambda\,r_{\rm
vir}\ . \eeq

The observed distribution of disk sizes suggests that the specific
angular momentum of the disk is similar to that of the halo (see DSS
and MMW), and so we assume $j_d/m_d=1$.  The distribution of disk
sizes is then determined\footnote{For a halo of a given mass and
redshift, we determine $r_{\rm vir}$ using NFW and eq.\ (6) of Bryan
\& Norman (1998); see also \S \ref{Vc}\,.} by the PS halo abundance
and by the distribution of spin parameters. The latter approximately
follows a lognormal distribution, \beq p(\lambda) d\lambda= \frac{1}
{\sigma_{\lambda} \sqrt{2 \pi}} \exp \left [-\frac{ \ln^2(\lambda/
\bar{\lambda})}{2\sigma_{\lambda}^2} \right] \frac{d\lambda}{\lambda}\
, \eeq with $\bar{\lambda}=0.05$ and $\sigma_{\lambda}=0.5$ following
MMW, who determined these values based on the N-body simulations of
Warren et al.\ (1992). Unlike MMW, we do not include a lower cutoff on
$\lambda$ due to disk instability.  If a dense bulge exists, it can
prevent bar instabilities, or if a bar forms it may be weakened or
destroyed when a bulge subsequently forms (Sellwood \& Moore 1999).

The distribution of disks is truncated at the low-mass end due to the
fact that gas pressure inhibits baryon collapse and disk formation in
shallow potential wells, i.e.\ in halos with a low circular velocity
$V_c$. In particular, photo-ionization heating by the cosmic UV
background heats the intergalactic gas to a characteristic temperature
of $\sim 10^{4-5}~{\rm K}$ and prevents it from settling into systems
with a lower virial temperature. Using a spherical collapse code,
Thoul \& Weinberg (1996) found a reduction of $\sim50\%$ in the
collapsed gas mass due to heating, for a halo of $V_c=50~{\rm
km~s}^{-1}$ at $z=2$, and a complete suppression of infall below
$V_c=30~{\rm km~s}^{-1}$. Three-dimensional numerical simulations
(Quinn, Katz, \& Efstathiou 1996; Weinberg, Hernquist, \& Katz 1997;
Navarro \& Steinmetz 1997) found a suppression of gas infall into even
larger halos with $V_c \sim 75~{\rm km~s}^{-1}$. We adopt a typical
cutoff value $V_{\rm cut}=50~{\rm km~s}^{-1}$ in the PS halo function,
requiring $V_c > V_{\rm cut}$ for the formation of disks.  We note,
however, that the appropriate $V_{\rm cut}$ could be lower at both
very low and very high redshifts when the cosmic UV background was
weak. In particular, the decline of the UV background at $z\sim 1$
allowed gas to condense in halos down to $V_c \sim 25~{\rm km~s}^{-1}$
(Kepner, Babul, \& Spergel 1997). Similarly, gaseous halos that had
formed prior to reionization, when the cosmic UV background had been
negligible, could have survived photo-ionization heating at later
times as long as they satisfied $V_c \ga 13~{\rm km~s}^{-1}$ (Barkana
\& Loeb 1999).

Aside from its relevance to lensing studies, the distribution of disk
sizes is useful for assessing the level of overlap of sources on the
sky, namely the confusion noise.  We first compute the geometric
optical depth of galactic disks, i.e., the fraction of the sky covered
by galactic disks. This corresponds to the probability of encountering
a galactic disk (within one exponential scale length) inside an
infinitesimal aperture.  Averaging over all random orientations, a
circular disk of radius $R_d$ at redshift $z_S$ occupies an angular
area of $2 (R_d/D_{OS})^2$.  The total optical depth then depends on
$V_{\rm cut}$. For $\Lambda$CDM with $V_{\rm cut}=50~{\rm km~s}^{-1}$,
we find the geometric optical depth to be $2.0 \times 10^{-4}$ when
integrated over all $z>10$ sources, $5.5 \times 10^{-3}$ for $z>5$
sources, $1.7\%$ for $z>3$ sources, $4.6\%$ for $z>1$ sources, and
$6.8\%$ for sources at all redshifts. If we lower $V_{\rm cut}$ to
$30~{\rm km~s}^{-1}$, the optical depth becomes $8.8 \times 10^{-4}$
for $z>10$ sources, $3.5\%$ for $z>3$ sources, and $11.3\%$ for all
source redshifts.

A more realistic estimate of confusion noise must include the finite
resolution of the instrument as well as its detection limit for faint
sources. We characterize the instrument's resolution by a minimum
circular aperture of angular diameter $\theta_a$. We include as
sources only those galactic disks which are brighter than some
threshold.  This threshold is dictated by $F_{\nu}^{\rm ps}$, the
minimum spectral flux\footnote{Note that $F_{\nu}^{\rm ps}$ is the
total spectral flux of the source, not just the portion contained
within the aperture.} required to detect a point-like source (i.e., a
source which is much smaller than $\theta_a$). For an extended source
of diameter $\theta_s \gg \theta_a$, we assume that the
signal-to-noise ratio can be improved by using a larger aperture, with
diameter $\theta_s$. The noise amplitude scales as the square root of
the number of noise (sky) photons, or the square root of the
corresponding sky area. Thus, the total flux needed for detection of
an extended source at a given signal-to-noise threshold is larger than
$F_{\nu}^{\rm ps}$ by a factor of $\theta_s/ \theta_a$.  We adopt a
simple interpolation formula between the regimes of point-like and
extended sources, and assume that a source is detectable if its flux
is at least $\sqrt{1+ (\theta_s/ \theta_a)^2}\, F_{\nu}^{\rm ps}$.

We can now compute the ``\,intersection probability'', namely the
probability of encountering a galactic disk (within one exponential scale
length) anywhere inside the aperture of diameter $\theta_a$. A face-on
circular disk of diameter $\theta_s=2 R_d/D_{OS}$ will overlap the aperture
if its center lies within a radius of $(\theta_a + \theta_s)/2$ about the
center of the aperture. Assuming a random orientation of the disk, the
average cross-section is then $\pi \theta_a^2/4 +1.323\ \theta_a \theta_s +
\theta_s^2/2$. We integrate this cross-section over the spin parameter
distribution and over the abundance of halos at all masses and redshifts.
The resulting intersection probability is closely related to the confusion
noise. If this probability is small then individual sources are resolved
from each other, since the aperture typically contains at most a single
detectable source.  We can also obtain a limit on the confusion noise from
sources below the flux detection threshold, by computing the same
intersection probability but including sources at all fluxes.

The flux $F_{\nu}$ of a given disk depends on its mass-to-light ratio,
which in turn depends on its star formation history and stellar mass
function. We adopt a semi-analytic starburst model similar to that of
Haiman \& Loeb (1998b), but different in detail. We assume that each
halo of mass $M$ hosts a disk of mass $m_d M$, of which a fraction
$f_d$ participates in star formation. Adopting a cosmological baryon
density of $\Omega_b h^2=0.02$, we define the star formation
efficiency $\eta$ so that $f_d m_d=\eta (\Omega_b/\Omega_0)$. We
assume a fixed universal value of $\eta$, and illustrate our results
for a high efficiency of $\eta=20\%$ (assumed unless indicated
otherwise) and for a low efficiency $\eta=2\%$. These values cover the
range of efficiencies suggested by observations of the metallicity of
the Ly$\alpha$ forest at $z=3$ (Haiman \& Loeb 1998b) and the
cumulative mass density of stars in the Universe at present (Fukugita,
Hogan, \& Peebles 1998). Note that $\eta=20\%$ and a particular value
of $F_{\nu}^{\rm ps}$ are equivalent to $\eta=2\%$ and a tenfold
decrease in $F_{\nu}^{\rm ps}$.

In order to determine the mass-to-light ratio of a halo of mass $M$ at
a redshift $z$, we assume that the mass $\eta (\Omega_b/\Omega_0) M$
is distributed in stars with a Salpeter mass function ($dN \propto
m^{-\alpha} dm$ with $\alpha$=2.35) from 1 $M_{\sun}$ up to 100
$M_{\sun}$. If the mass function were extended to masses below 1
$M_{\sun}$, the additional stars would contribute significant mass but
little luminosity, so this would essentially be equivalent to a
reduction in $\eta$. We use the stellar population code of Sternberg
(1998) with Z=0.001 stellar tracks and Z=0.006 stellar spectra. We
assume that the age of the stellar population equals that of the dark
matter halo, whose age is determined from its merger history. The
formation redshift $z_{\rm \,form}>z$ is defined as the time at which
half the mass of the halo was first contained in progenitors more
massive than a fraction $f$ of $M$.  We set $f=0.5$ and estimate the
formation redshift (and thus the age) using the extended
Press-Schechter formalism (see, e.g., Lacey \& Cole 1993). At high
redshifts, the young age of the Universe and high halo merger rate
imply young stellar populations which are especially bright at
rest-frame UV wavelengths. At each redshift $z$ we calculate the halo
spectral flux by averaging the composite stellar spectrum over the
wavelengths corresponding to the observed {\it NGST}\, spectral range
of 0.6--3.5$\mu$m. We also include a Ly$\alpha$ cutoff in the spectrum
due to absorption by the dense Ly$\alpha$ forest at all redshifts up
to that of the source. We do not, however, include dust
extinction. Despite the generally low metallicity at high redshifts,
extinction could be significant since observations correspond to
rest-frame UV wavelengths (Loeb \& Haiman 1997).

Our starburst model is expected to describe galaxies at high
redshifts, but it may fail at redshifts $z\la 2$. The model relies on
two key assumptions, namely that stars form in disks, and that the
stars in each galaxy have formed since the last major merger of its
halo. At high redshifts, the fraction of gas that has collapsed into
halos is small, and the fraction that has turned into stars is even
smaller. Thus, a high-redshift galaxy is expected to be gas-rich
whether it forms in a merger or accretes most of its gas from the
intergalactic medium. Such a galaxy is likely to form most of its
stars in a disk after the gas cools and settles onto a plane. At low
redshifts, on the other hand, disk galaxies may have converted most of
their gas into stars by the time they merge. In this case, the merger
may form a massive elliptical galaxy rather than a disk-dominated
galaxy. Indeed, elaborate semi-analytic models indicate that the stars
in elliptical galaxies are typically much older than their halo merger
age (e.g., Thomas \& Kauffmann 1999), in agreement with the red colors
of ellipticals which suggest old stellar populations.  Although the
increased presence of elliptical galaxies invalidates our model for
the mass-to-light ratios of galaxies at low redshifts, our results for
the size distribution of galaxies may remain approximately valid.
Theoretical considerations based on the virial theorem, as well as
numerical simulations, suggest that the characteristic size of a
galactic merger remnant is smaller by a factor of $\la 1.5$ than the
size expected for a disk galaxy of the same mass and velocity
dispersion (Hausman \& Ostriker 1978; Hernquist et al. 1993).

\subsection{Numerical Results}

Figure \ref{figconf} shows the total intersection probability as a
function of limiting flux (right panel), for all sources with $z>0$,
$z>2$, $z>5$, and $z>10$, from top to bottom. The total probability is
dominated by the contribution of sources at low redshifts, which is
relatively insensitive to the limiting flux (or to $\eta$). All curves
assume the $\Lambda$CDM model with a circular-velocity cutoff for the
host halo of $V_{\rm cut}=50~{\rm km~s}^{-1}$. The aperture diameter
is chosen to be $\theta_a=0\farcs 06$, close to the expected {\it
NGST}\, resolution at $2\mu$m. With $F_{\nu}^{\rm ps}=1$
nJy, the total intersection
probability for all redshifts is $8.9\%$ (or $5.6\%$ if $\eta=2\%$) in
$\Lambda$CDM (and it is $10\%$ or less also in the SCDM and OCDM
models). The probability increases to $15\%$ ($6.2\%$ if $\eta=2\%$)
if $V_{\rm cut}=30~{\rm km~s}^{-1}$ instead of $50~{\rm km~ s}^{-1}$.
The contribution from sources at $z>5$ is $1.0\%$ ($9.0\times 10^{-4}$
if $\eta=2\%$). Thus the chance for overlapping sources will be small
for {\it NGST}.\, If the resolution were $\theta_a = 0\farcs 12$, the
probability would be $12\%$ ($6.6\%$ if $\eta=2\%$) for all redshifts
and $1.7\%$ ($0.14\%$ if $\eta=2\%$) for sources at $z>5$. If we
include all sources regardless of flux then the probability becomes
independent of $\eta$, and (with $\theta_a = 0\farcs 06$) it equals
$9.1\%$ if $V_{\rm cut}=50~{\rm km~ s}^{-1}$ and $18.8\%$ if $V_{\rm
cut}=30~{\rm km~s}^{-1}$. The contribution from sources below the
detection threshold is small due to the $V_c$ cutoff, i.e.\ the fact
that the photo-ionizing background prevents the formation of galaxies
in small dark matter halos. This fact should eventually result in a
turnover, where the number counts do not increase with decreasing
flux. However, the turnover occurs somewhat below 1 nJy, a flux that
is much smaller than the detection threshold of current observations
such as the Hubble Deep Field.

In summary, we have shown that confusion noise for {\it NGST}\, will
be low, assuming that there is one galaxy per halo and that the
luminous stars form primarily in disks.  Note that we have not
included the possible confusion noise from multiple galaxies per halo,
from clustered or interacting galaxies, or from galaxies being
observed as separate fragments rather than smooth disks. We also have
not included the confusion noise from stars and other sources in our
own galaxy. Also note that with no flux limit on sources, the
intersection probability approaches unity only if the aperture is
increased to $0\farcs 9$.

Our model predicts the size distribution of galaxies at various
redshifts. Figure \ref{figsize} shows the fraction of the total number
counts contributed by sources with diameters greater than $\theta$, as
a function of $\theta$. The size distributions are shown for a high
efficiency ($\eta=20\%$, solid curves) and for a low efficiency
($\eta=2\%$, dotted curves) of star formation. Each curve is marked by
the lower limit of the corresponding redshift range, with `0'
indicating sources with $0<z<2$, and similarly for $2<z<5$, $5<z<10$,
and $z>10$. All curves include a cutoff of $V_{\rm cut}=50~{\rm
km~s}^{-1}$ and a limiting point source flux of 1 nJy, and all are for
the $\Lambda$CDM model. The vertical dashed line in Figure
\ref{figsize} indicates the {\it NGST}\, resolution of $0\farcs
06$. Note that increasing $\eta$ leads to a decrease in the typical
angular size of galaxies, since the set of observable galaxies then
includes galaxies which are less massive, and thus generally
smaller. However, a tenfold increase in $\eta$ decreases the observed
angular sizes of $z>10$ galaxies by only a factor of two.

The typical observed size of faint disks (i.e., of all disks down to 1
nJy) is $0\farcs4$ for sources at $0<z<2$, $0\farcs2$ for sources at
$2<z<5$, $0\farcs10$ (or $0\farcs15$ if $\eta=2\%$) for sources at
$5<z<10$, and $0\farcs065$ (or $0\farcs11$ if $\eta=2\%$) for sources
at $z>10$. Roughly $60\%$ of all $z>10$ sources (or $90\%$ if
$\eta=2\%$) can be resolved by {\it NGST},\, and the fraction is at
least $85\%$ among lower redshift sources. Thus, the high resolution
of {\it NGST}\, should make most of the detected sources useful for
weak lensing. If reliable shape measurements require a diameter equal
to twice the resolution scale (probably overly pessimistic), then the
useful ($\theta > 0\farcs 12$) fractions are $13\%$ for $z>10$, $40\%$
for $5<z<10$, and $80\%$ for $2<z<5$ sources. If $\eta=2\%$, the
corresponding fractions are $40\%$ for $z>10$, $65\%$ for $5<z<10$,
and $80\%$ for $2<z<5$. These results are all in the $\Lambda$CDM
model, but disk sizes in the SCDM and OCDM models differ by only about
$10\%$.

As noted by Schneider \& Kneib (1998), ground-based telescopes that
are not equipped with adaptive optics or interferometry would be
unable to resolve most of the high-redshift sources, even if they
could reach the same flux sensitivity as {\it NGST}.\, For example, a
ground-based survey down to 1 nJy with, e.g., $0\farcs 75$ seeing at
$2\mu$m, could resolve only $0.003\%$ of the $z>10$ sources, with
corresponding fractions of $0.1\%$ for $5<z<10$, and $2\%$ for
$2<z<5$. If $\eta=2\%$ then the resolved fractions are $0.03\%$ for
$z>10$, $0.8\%$ for $5<z<10$, and $4\%$ for $2<z<5$. Thus, the high
resolution of {\it NGST}\, is crucial for resolving faint galaxies at the
redshifts of interest.

Current observations of galaxy sizes at $z>2$ are inadequate for a detailed
comparison with our models. Gardner \& Satyapal (1999, in preparation) have
determined the sizes of galaxies in the Hubble Deep Field South, finding
typical half-light radii of $0\farcs 1$ with a very large scatter. This
sample likely includes a wide range of redshifts, and it is expected to be
strongly biased toward small galaxy sizes. Given the steep luminosity
function of the detected galaxies, most of them are detected very close to
the detection limit, especially those at high redshift. Of course, galaxies
near the flux threshold can be detected only if they are nearly point
sources, while large galaxies are excluded from the sample because of their
low surface brightness.

Since most galaxies will be resolved by {\it NGST},\, predictions for
the total number counts are affected by the higher flux needed for the
detection of extended objects relative to point sources. For a point
source flux limit of 1 nJy and $\eta=20\%$, the total number counts
are reduced (relative to a size-independent flux limit of 1 nJy)
by a factor of 2 for $z>10$ and by only $10\%$ for $5<z<10$. The
reduction for $z<10$ sources is small if $\eta=20\%$, since in this
case the total flux of most $z<10$ sources is greater than 1 nJy, 
and these galaxies can still be detected even as extended
objects. However, the reduction in number counts is more significant
if $\eta=2\%$, with a factor of 8 for $z>10$, 4 for $5<z<10$, and 2.5
for $2<z<5$.

\begin{figure}
\epsscale{0.7}
\plotone{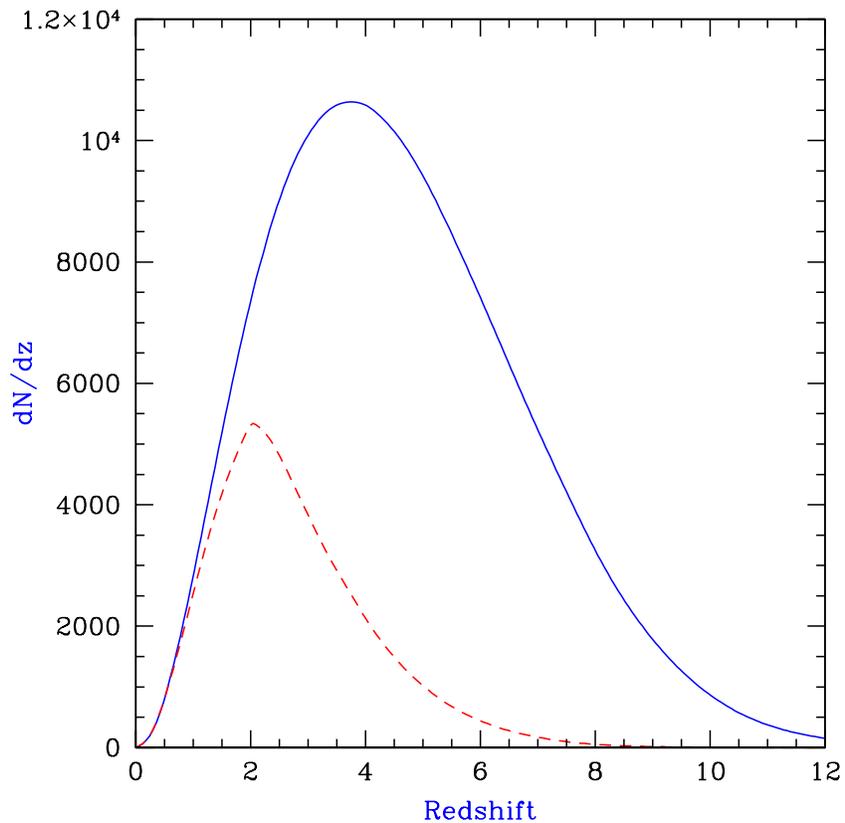}
\caption{Predicted redshift distribution of galaxies observed with
{\it NGST}.\, The distribution in the $\Lambda$CDM model is shown for
a high efficiency ($\eta=20\%$, solid curve) or a low efficiency
($\eta=2\%$, dashed curve) of star formation. The plotted quantity is
$dN/dz$, where $N$ is the number of galaxies per {\it NGST}\, field of
view. All curves assume a limiting point source flux of 1 nJy.}
\label{figzSource}
\end{figure}

We show in Figure \ref{figzSource} the resulting prediction for the
redshift distribution of the galaxy population observed with {\it
NGST}.\, We assume the $\Lambda$CDM model and plot $dN/dz$, where $N$
is the number of galaxies per {\it NGST}\, field of view. The solid
curve assumes a high efficiency ($\eta=20\%$) of star formation and
the dashed curve assumes a low efficiency ($\eta=2\%$). All curves
assume a limiting point source flux of 1 nJy. The total number per
field of view of galaxies at all redshifts is $N=59,000$ for
$\eta=20\%$ and $N=15,000$ for $\eta=2\%$. The fraction of galaxies
above redshift 5 is sensitive to the value of $\eta$ -- it equals
$40\%$ for $\eta=20\%$ and $7.4\%$ for $\eta=2\%$ -- but the number of
$z>5$ galaxies is large ($\sim 1000$) even for the low efficiency.
The number of $z>5$ galaxies predicted in SCDM is close to that
in $\Lambda$CDM, but in OCDM there are twice as many $z>5$ galaxies.

\subsection{The Surface Brightness of Lensed Sources}

In our estimates of the lensing rate in \S 2 we implicitly made two
important assumptions: (i) the source is smaller than the image separation,
so that the two images of the source are not blended; (ii) the surface
brightness of the background source is comparable to or higher than that of
the foreground lens, otherwise the background source could not be detected
when it is superimposed on the lens galaxy.  These assumptions are
trivially justified for the point-like images of quasars. In the context of
galactic sources, we can apply our estimates of disk sizes to test these
two assumptions quantitatively.

A lensed galaxy is generally much smaller than the separation of its
two images. The combination of Figures \ref{figang} and \ref{figsize}
shows that, regardless of the source redshift, the typical image
separation is at least four times as large as the typical diameter of
a source galaxy detected by {\it NGST}.\, Thus, the majority of all
lensed sources will not be blended. Note, however, that if ellipticity
or shear are included then some of the resulting four-image systems
may include arcs produced by several blended images.

In order to compare the surface brightness of source galaxies to that
of lens galaxies, we calculate the redshift evolution of the mean
surface brightness of galaxies. At high redshifts, we may apply our
disk starburst model to find the surface brightness of a galaxy from
the predicted size, mass, and mass-to-light ratio of its disk. We
compute the average surface brightness (as observed in the {\it
NGST}\, spectral range) out to one exponential scale length. Figure
\ref{figSB} shows this surface brightness $\mu$ (expressed in nJy per
square arcsecond) averaged over all galaxies at each redshift, in the
$\Lambda$CDM model only (as the OCDM and SCDM models yield very
similar results). Solid lines show the mean at $z>2$, where galaxies
are weighed by their number density and their mass-to-light ratios are
derived from the starburst model. As discussed at the end of \S
\ref{limits}, although our model for the size distribution of galaxies
should remain approximately valid at low redshifts, the starburst
model may fail to predict the correct mass-to-light ratio of the
stellar population at $z\la 2$, particularly for the lens galaxies. These
lenses tend to be massive elliptical galaxies, with stellar
populations that may be much older than the merger timescale assumed
in our starburst model. In order to estimate the surface brightness of
lens galaxies, we adopt a simple alternative model in which all their
stars are uniformly old. The dashed lines in Figure \ref{figSB} show
(for $z<2$) the mean surface brightness of lensing galaxies (i.e.,
where galaxies are weighed by the product of their number density and
their lensing cross-section), assuming that their stars formed at
$z=5$. In each case (i.e., for source galaxies or for lens galaxies),
the upper curve assumes a high efficiency ($\eta=20\%$) and the lower
curve assumes a low efficiency ($\eta=2\%$) of incorporating baryons
into stars in the associated halos. All curves include a cutoff
velocity of $V_{\rm cut}=50~{\rm km~s}^{-1}$ and a limiting point
source flux of 1 nJy.

As is apparent from Figure \ref{figSB}, the mean surface brightness of
galaxies varies, for a fixed $\eta$, by a factor of $\la 2$ over all
redshifts above 2, despite the large range in luminosity distances
from the observer. Several different factors combine to keep the
surface brightness nearly constant. Except for redshift factors, the
surface brightness is proportional to the luminosity over the square
of the disk radius, and the luminosity is in turn equal to the disk
mass divided by its mass-to-light ratio. Although the typical mass of
halos decreases at high redshifts, two other effects tend to increase
the surface brightness. First, high redshift disks are compact due to
the increased mean density of the Universe. The second effect results
from the low mass-to-light ratio of the young stellar populations of
high redshift disks, which makes these galaxies highly luminous
despite their small masses. For example, the mean ratio of halo mass
to disk luminosity for $z=2$ galaxies (with $\eta=20\%$ and
$F_{\nu}^{\rm ps}=1$ nJy) is 14 in solar units, and this decreases to
3.8 at $z=5$ and 1.2 at $z=10$.  This evolution in the mass-to-light
ratio includes the so-called K-correction, i.e.\ the fact that for
higher-redshift sources the {\it NGST}\, filter corresponds to shorter
rest-frame wavelengths.

Acting alone, the factors discussed above would result in a sharp
increase with redshift in the surface brightness of galaxies.  Additional
redshift effects, however, counter-balance these other factors. According to
the Tolman surface brightness law, the expansion of the universe yields a
factor of $(1+z)^{-4}$ regardless of the values of the cosmological
parameters. This redshift factor dominates and produces an overall decrease
in $\mu$ among lens galaxies at low redshifts (up to $z \sim 1.5$). At
these low redshifts, all galaxies are detected regardless of $\eta$, so the
overall $\mu$ is exactly proportional to $\eta$. At higher redshifts, the 1
nJy flux limit preferentially removes low surface brightness galaxies from
the detected sample. The resulting bias toward high surface brightness is
larger if $\eta=2\%$, and this decreases the difference in $\mu$ between
the cases of $\eta=2\%$ and $\eta=20\%$. The mass-to-light ratio begins to
decrease rapidly at $z \ga 1.5$, and at $z>2$ the various factors combine
to produce a slow variation in $\mu$.

Although there is only a modest redshift evolution in the surface
brightness of galaxies, there is an additional difficulty in detecting
lensed sources. Lensing galaxies are biased toward larger circular
velocities, i.e.\ toward larger masses at each redshift. Since a
galaxy that is more massive is usually also more luminous, its surface
brightness tends to be larger. As shown in Figure \ref{figSB}, this
tendency makes the mean surface brightness of lenses somewhat higher
than that of sources, despite our assumption of an old stellar
population in lens galaxies. Consider, for example, a source at
redshift 5 which is multiply imaged. The mean lens redshift for
$z_S=5$ is $z_L=1.4$. If we select the source from the general galaxy
population and the lens from the population of lenses, then the
typical source-to-lens surface brightness ratio is 1:3 if $\eta=20\%$
(or close to 1:1 if $\eta=2\%$).

Even though lens galaxies might have a somewhat higher mean surface
brightness than the sources which they lens, it should be possible to
detect lensed sources since (i) the image center will typically be
some distance from the lens center, of order half the image
separation, and (ii) the younger stellar population and higher
redshift of the source will make its colors different from those of
the lens galaxy, permitting an easy separation of the two in
multi-color observations.  These two helpful features, along with the
source being much smaller than the lens and the image separation, are
evident in the currently known systems which feature galaxy-galaxy
lensing. These include two four-image `Einstein cross' gravitational
lenses discovered by Ratnatunga et al.\ (1995) in the Groth-Westphal
strip, and a lensed three-image arc detected in the Hubble Deep Field
South and studied in detail by Barkana et al.\ (1999). In these cases
of moderate redshifts and optical/UV observations, the sources appear
bluer than the lens galaxies.

In the infrared range of {\it NGST},\, high-redshift sources are
expected to generally be redder than their low redshift lenses, since
the overall redshift has a dominant effect on the spectrum. Suppose,
e.g., that $z_S=5$ and $z_L=1.4$. We divide the {\it NGST}\, spectral
range into four logarithmically-spaced parts (in order of increasing
frequency). For a given spectrum, we find the fraction of the total
luminosity which is emitted in each frequency quadrant. The mean
fractions for $z_S=5$ galaxies are 0.37, 0.21, 0.26, and 0.16,
respectively, while the fractions for $z_L=1.4$ lenses (assuming, as
above, that their stars formed at redshift 5) are 0.16, 0.29, 0.39,
and 0.16 . Thus, if we use the lowest frequency quadrant, the source
will be brighter than the lens by an additional factor of 2.3 relative
to the source-to-lens luminosity ratio when we use the full {\it
NGST}\, bandwidth. Note that we have not included here extinction,
which could further redden the colors of lensed sources.

\section{Conclusions}

We have calculated the lensing probability of high-redshift galaxies
or quasars by foreground dark matter halos. We found that the lensing
optical depth for multiple imaging of sources increases by a factor of
4--6 from $z_S=2$ to $z_S=10$. With a magnification bias of $\sim 5$
expected for $z_S>5$ sources, the fraction of apparent sources which
form one of the images of a lensed source reaches $\sim 5\%$ for
sources at $z_S=10$ (see Table 1).  Among lenses with image
separations below $5 \arcsec$, the typical image separation (in
$\Lambda$CDM) drops from $1\farcs1$ at $z_S=2$ to $0.5 \arcsec$ at
$z_S=10$. With its expected $\sim 0\farcs06$ resolution, {\it NGST}\,
can resolve $\sim 85\%$ of the lenses with $z_S=10$. Assuming the
number counts predicted by Haiman \& Loeb (1998b), the estimated
number of lensed sources above 1 nJy per field of view of {\it NGST}\,
is roughly 5 for $z>10$ quasars, 10 for $z>5$ quasars, 1--15 for
$z>10$ galaxies, and 30--200 for $z>5$ galaxies. Note that these
values are in a $\Lambda$CDM cosmology; the number of $z>10$ galaxies
is smaller by a factor of $\sim 3$ in SCDM but larger by a factor of
$\sim 10$ in OCDM.

Although only a small fraction of the sources are multiply imaged, all
sources are mildly altered by gravitational lensing due to
foreground objects. For a source that is not multiply imaged, the
cross-section for an amplification of at least $A$ varies as
$1/(A-1)^2$ for a SIS lens. Thus, for $z_S=10$ the optical depth is
unity for an amplification of $A=1.1$ or greater. This implies that
extended sources at high redshifts are significantly distorted due to
lensing. A typical $z=10$ source is magnified or de-magnified by 
$\sim 10\%$ and also has an ellipticity of at least $10\%$ due to
lensing.

\begin{figure}
\epsscale{0.7}
\plotone{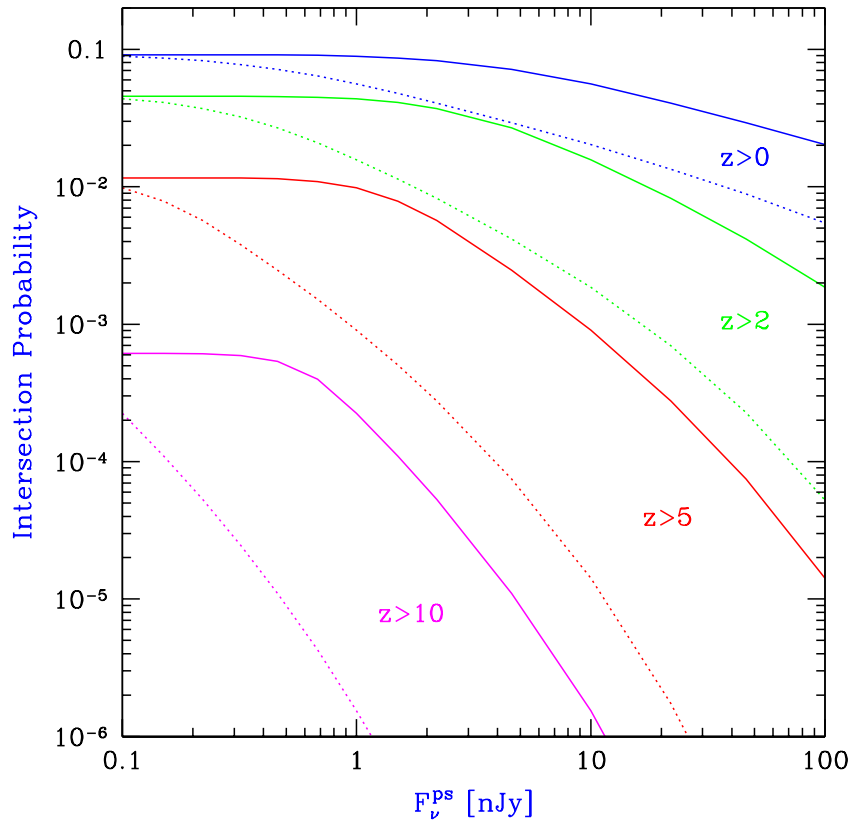}
\caption{Probability of encountering a galactic disk inside the {\it
NGST}\, aperture of diameter $0\farcs 06$. We show this intersection
probability as a function of limiting point source flux for a high
efficiency ($\eta=20\%$, solid curves) and for a low efficiency
($\eta=2\%$, dotted curves) of star formation. In each case, the curves
include all disks with $z>0$, $z>2$, $z>5$, and $z>10$, respectively, from
top to bottom.  We assume the $\Lambda$CDM model and a circular-velocity
cutoff for the host halo of $V_{\rm cut}=50\ {\rm km\ s}^{-1}$.}
\label{figconf}
\end{figure}

\begin{figure}
\epsscale{0.7}
\plotone{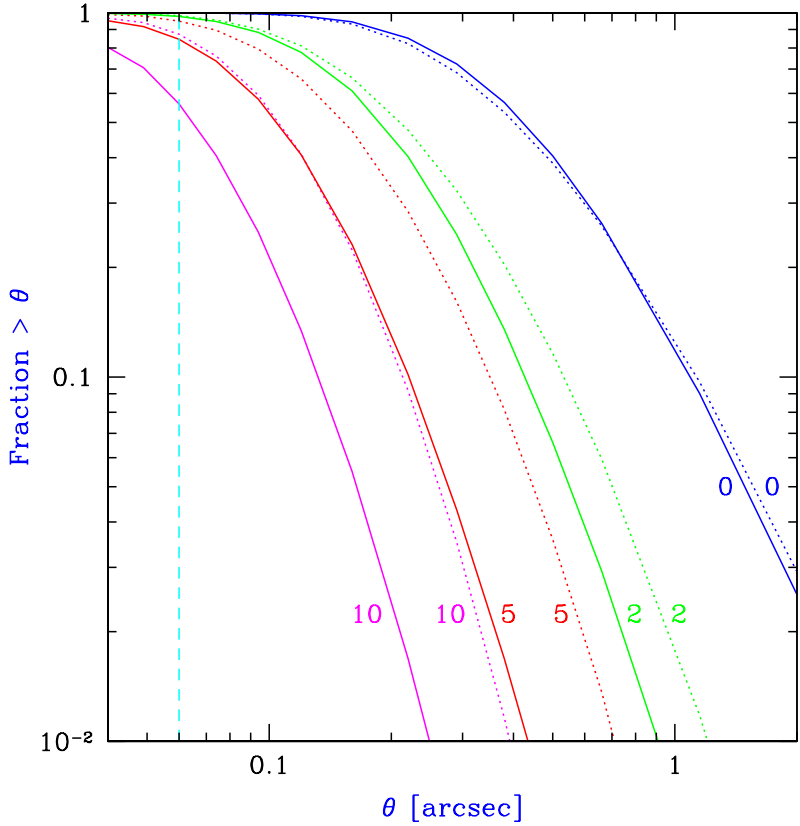}
\caption{Distribution of galactic disk sizes in various redshift intervals,
in the $\Lambda$CDM model. Given $\theta$ in arcseconds, each curve shows
the fraction of the total number counts contributed by sources larger than
$\theta$. The diameter $\theta$ is measured out to one exponential scale
length. We assume either a high efficiency ($\eta=20\%$, solid curves) or a
low efficiency ($\eta=2\%$, dotted curves) of star formation. We indicate
next to each curve the lower limit of the redshift interval. Thus `0'
indicates sources with $0<z<2$, and similarly we consider sources with
$2<z<5$, $5<z<10$, and $z>10$. All curves include a cutoff velocity of
$V_{\rm cut}=50\ {\rm km\ s}^{-1}$ and a limiting point source flux of 1
nJy. The vertical dashed line indicates the {\it NGST}\, resolution of
$0\farcs 06$.}
\label{figsize}
\end{figure}

We have also predicted the size distribution of galactic disks at high
redshifts (see Figure \ref{figsize}) and found that the angular
resolution of {\it NGST}\, will be sufficiently high to avoid
confusion noise due to overlapping sources. Indeed, with a 1 nJy flux
limit the probability of encountering a galactic disk inside an
aperture of $0\farcs06$ diameter is $8.9\%$ for $\Lambda$CDM, of which
$4\%$ comes from $z>2$ sources, $1\%$ comes from $z>5$ sources, and
only $0.02\%$ is contributed by $z>10$ sources (see Figure
\ref{figconf}). These values are for a high star formation efficiency of
$\eta=20\%$, and they are reduced if $\eta=2\%$.

\begin{figure}
\epsscale{0.7}
\plotone{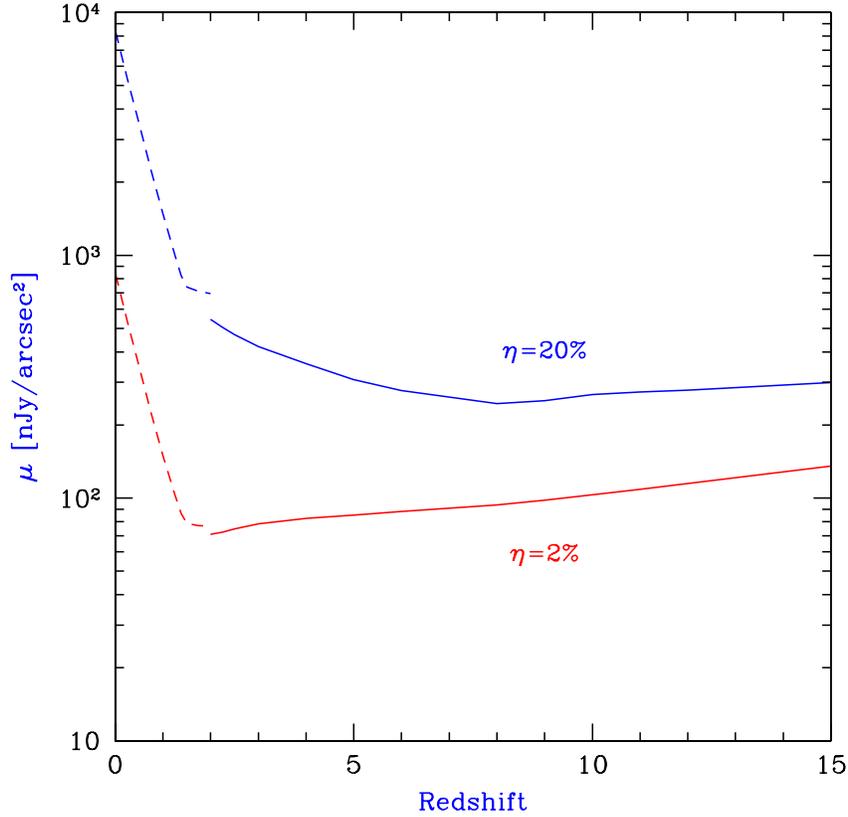}
\caption{Redshift evolution of the mean surface brightness $\mu$ of
galaxies, in the $\Lambda$CDM model, averaged over the {\it NGST}
wavelength band. 
In calculating this mean, the contribution of each disk is its average
surface brightness out to one exponential scale length. Solid lines show
(for $z>2$) the mean calculated for all galaxies weighed by their number
density, with the mass-to-light ratios derived from the starburst
model. Dashed lines show (for $z<2$) the mean for lens galaxies (i.e.,
weighed by the product of their number density and their lensing
cross-section), with the mass-to-light ratios derived from assuming that
all their stars formed at $z=5$. In each case, the upper curve assumes a
high efficiency ($\eta=20\%$) and the lower curve assumes a low efficiency
($\eta=2\%$) of converting gas into stars. All curves include a cutoff
velocity of $V_{\rm cut}=50\ {\rm km\ s}^{-1}$ and a limiting point source
flux of 1 nJy.}
\label{figSB}
\end{figure}

In our estimates of the lensing rate in \S 2, we assumed that a lensed
source can be detected even when its images overlap the lensing
galaxy. We showed in \S 3 that the mean surface brightness of galaxies
evolves modestly above redshift 2 (see Figure \ref{figSB}). Although
the surface brightness of a background source will typically be
somewhat lower than that of the foreground lens, the lensed images
should be detectable since they are offset from the lens center and
their colors are expected to differ from those of the lens galaxy.

Although the typical size of sources decreases with increasing
redshift, at least $60\%$ of the $z>10$ galaxies above 1 nJy can still
be resolved by {\it NGST}.\, This implies that the shapes of these
high redshift galaxies can be studied with {\it NGST}.\, We have also
found that the high resolution of {\it NGST}\, is crucial in making
the majority of sources on the sky useful for weak lensing studies.

When we assumed a 1 nJy flux limit for detecting point sources, we
included the fact that resolved sources require a higher flux in order
to be detected with the same signal-to-noise ratio. Therefore, estimates
of number counts that assume a constant flux limit of 1 nJy for all
sources overestimate the number counts by a factor of 2 for $z>10$
sources and a star formation efficiency of $\eta=20\%$, or by as much
as a factor of 8 if $\eta=2\%$. Even with this limitation, though,
{\it NGST}\, should detect a total (over all redshifts) of roughly one
galaxy per square arcsecond for $\eta=20\%$ (or one per 4 square
arcseconds if $\eta=2\%$).

In conclusion, the field of gravitational lensing is likely to benefit
greatly over the next decade from the combination of unprecedented
sensitivity and high angular resolution of {\it NGST}.\,

\acknowledgments

We thank Zoltan Haiman for providing number count data from earlier work.
We are also grateful to Tal Alexander and Amiel Sternberg for numerical
results of their stellar population model, and to David Hogg for useful
discussions. RB acknowledges support from Institute Funds. This work was
supported in part by NASA grants NAG 5-7039 and NAG 5-7768 for AL.


\begin{deluxetable}{lrrrr}
\tablecaption{Fraction of Lensed Sources}
\tablenum{1}
\tablehead{
\colhead{Cosmological Model} & \colhead{PS halos, $z_S=10$} & 
\colhead{PS halos, $z_S=5$}    & \colhead{PS halos, $z_S=2$} 
& \colhead{Galaxies, $z_S=2$} }
\startdata
$\Lambda$CDM & $5.5 \%$ & $3.3 \%$
& $1.0\%$ & $0.80 \%$ \nl
OCDM & $4.5 \%$ & $2.5 \%$ &
$0.70 \%$ & $0.38\%$ \nl
SCDM & $5.5 \%$ & $3.4\%$ &
$1.3\%$ & $0.26\%$ \nl
\enddata
\tablecomments{In all cases we have assumed a factor of 5 from 
magnification bias.}
\end{deluxetable}

\end{document}